\documentclass[prc,aps,nofootinbib,superscriptaddress,preprintnumbers]{revtex4}
\usepackage{graphicx}
\usepackage{amsmath}
\usepackage{amsfonts,amsbsy}
\usepackage{amssymb}

\def\empile#1\over#2{\mathrel{\mathop{\kern 0pt#1}\limits_{#2}}}

\def\lsim{ \,\, \vcenter{\hbox{$\buildrel{\displaystyle <}\over\sim$}}
 \,\,}
\newcommand{\be}{\begin{equation}}
\newcommand{\ee}{\end{equation}}

\begin{document}

\title{KNO scaling of fluctuations in $pp$ and $pA$, and
eccentricities in heavy-ion collisions}
\author{Adrian Dumitru}
\affiliation{RIKEN BNL Research Center, Brookhaven National Laboratory,
Upton, NY-11973, USA
}
\affiliation{Department of Natural Sciences, Baruch College,
CUNY, 17 Lexington Avenue, New York, NY 10010, USA
}
\author{Yasushi Nara}
\affiliation{Akita International University, Yuwa, Akita-city
  010-1292, Japan 
}

\begin{abstract}
Multiplicity fluctuations at midrapidity in $pp$ collisions at high
energies are described by a negative binomial distribution and exhibit
approximate Koba-Nielsen-Olesen (KNO) scaling. We find that these KNO
fluctuations are important also for reproducing the multiplicity
distribution in $d+Au$ collisions observed at RHIC, adding to the
Glauber fluctuations of the number of binary collisions or
participants. We predict that the multiplicity distribution in $p+Pb$
collisions at the LHC also deviates little from the KNO scaling
function. Finally, we analyze various moments of the eccentricity of
the collision zone in $A+A$ collisions at RHIC and LHC and
find that particle production fluctuations increase fluctuation
dominated moments such as the triangularity $\epsilon_3$
substantially.
\end{abstract}

\preprint{RBRC 941}

\maketitle

Charged particle multiplicity distributions in the central region of
inelastic (non-single diffractive) $\bar{p}+p$ collisions at high
energies were shown by the UA1 and UA5 collaborations to follow a
negative binomial distribution (NBD)~\cite{UA1,UA5} exhibiting
approximate ``KNO scaling''~\cite{Koba:1972ng} over at least a limited
range of multiplicities excluding the tails. Within the framework of
high-energy QCD they may be thought to arise from fluctuations of the
density of large-$x$ valence charges~\cite{Gelis:2009wh} and of
stochastic emissions in the rapidity evolution
ladders~\cite{Mueller:1996te,Avsar:2010rf} leading from the rapidity
of the sources to the central region.

Collisions of hadrons or heavy ions at high energies release a large
number of gluons from their wave functions. In fact, the wave
function of a hadron boosted to (nearly) the light cone is so densely
packed with gluons that they may ``overlap'', leading to non-linear
interactions~\cite{GLR}. Therefore, at high energies the colliding
hadrons can be treated as a high occupancy gluon field. This dense
system is nowadays referred to as Color Glass Condensate
(CGC)~\cite{Gelis:2010nm}.

Here, we use the ``$k_\perp$-factorization'' approach~\cite{GLR} to
compute particle production in high-energy collisions:
\be
\left< \frac{dN^{A+B\to g}}{dy \,d^2{\bf r}_\perp}\right> = K\,
\frac{N_c}{N_c^2-1}
\int\frac{d^2{\bf p}_\perp}{p_\perp^2} \int\limits^{p_\perp}{d^2{\bf k}_\perp}\, 
\alpha_s(Q)\;
\Phi\left(\frac{|{\bf p}_\perp+{\bf k}_\perp|}{2},x_1\right)\,
\Phi\left(\frac{|{\bf p}_\perp-{\bf k}_\perp|}{2},x_2\right)~,
\label{eq:kt}
\ee
where $N_c=3$ is the number of colors in QCD and $K\simeq1.5-2$ is a
multiplicative factor to account for corrections to this LO
formula. Further, we assume that the hadron multiplicity is
proportional to the multiplicity of gluons. These factors may depend
somewhat on the initial condition for small-$x$ evolution of
$\Phi(k_\perp)$ but were found to be approximately independent of
collision energy or centrality (for heavy-ion
collisions)~\cite{AlbDum}. Also, in eq.~(\ref{eq:kt})
$x_{1,2}=(p_\perp/\sqrt{s})\; \exp \pm y$ and the strong coupling is
evaluated at the scale $Q=\mathrm{max}(|{\bf p}_\perp+{\bf k}_\perp| ,
|{\bf p}_\perp-{\bf k}_\perp|)/2$.

We require the evolution of the so-called unintegrated gluon
distribution $\Phi(k_\perp,x)$ (per unit transverse area) 
with the light-cone momentum fraction
$x$, starting from an initial condition at $x_0\simeq10^{-2}$. This
is obtained by solving the non-linear Balitsky-Kovchegov (BK)
equation~\cite{BK} with the running-coupling kernel according to
Balitsky's prescription~\cite{AK}. Specifically, we use the
unintegrated gluon distribution ``set MV'' from ref.~\cite{AlbDum}.

For the case of heavy-ion projectiles and/or targets, we allow for
fluctuations of the locations of the sources (i.e., of the valence
charges at $x_0$) for the small-$x$ fields in the transverse plane
before the collision~\cite{AlbDum,DrescherNara}.  This leads to
fluctuations of the ``geometry'' of the collision zone from
configuration to configuration, and to fluctuations of the number of
participants $N_{\rm part}$ and the number of collisions $N_{\rm
  coll}$ which are determined within the well-known Glauber
approach. Note that eq.~(\ref{eq:kt}) refers to a {\em single} such
configuration. We computed these ``geometry'' fluctuations assuming
that the {\em hard valence charges} are smeared over a finite and
energy independent area $\sigma_0 \sim \sigma_{NN}(200~{\rm GeV}) =
4.2$~fm$^2$. This reduces higher-order eccentricities as compared to
point-like nucleons~\cite{Nara:2011ee} which are used in some
Monte-Carlo Glauber simulations. Our numerical simulations do not
account for correlations (in the transverse plane) among the valence
charges which could further suppress geometry
fluctuations~\cite{correlGlauber}.

The unintegrated gluon densities $\Phi(k_\perp,x)$ from
eq.~(\ref{eq:kt}) have already been averaged over the local
fluctuations of the valence charges {\em in color space}, and over the
evolution ladders. It is in this sense that we interpret
eq.~(\ref{eq:kt}) as a mean (local) multiplicity. In each cell
$\Delta^2 {\bf r}_\perp$ of the transverse plane the actual
multiplicity is a NBD random variable,
\be \label{eq:NBD}
P(n) = \frac{\Gamma(k+n)}{\Gamma(k) \, \Gamma(n+1)}
\frac{\bar n^n k^k}{(\bar n + k)^{n+k}}~.
\ee
Here, $\bar n\equiv \langle dN/d\eta\,d^2{\bf r}_\perp\rangle\,
\Delta^2 {\bf r}_\perp \Delta\eta$ is the mean multiplicity from
eq.~(\ref{eq:kt}) in a given cell and $k$ is the fluctuation
parameter: smaller $k$ correspond to larger fluctuations about the
mean and KNO scaling is obtained when $k\ll\bar n$ (see below). We
finally average over geometric configurations of sources in the ${\bf
  r}_\perp$ plane described above, and over the impact parameter of
the collision.

There have been numerous theoretical discussions of multiplicity
fluctuations in high-energy collisions. Ref.~\cite{Gelis:2009wh}, in
particular, argued that NBD multiplicity fluctuations arise in a
semi-classical calculation of gluon production from dense valence
charge sources. They obtain that the fluctuation parameter $k$
is proportional to the density per unit transverse area of
valence charge squared, i.e.\ to the saturation momentum $Q_s^2$ at
$x_0$.

\begin{figure}[htb]
\begin{center}
\includegraphics[width=0.32\textwidth]{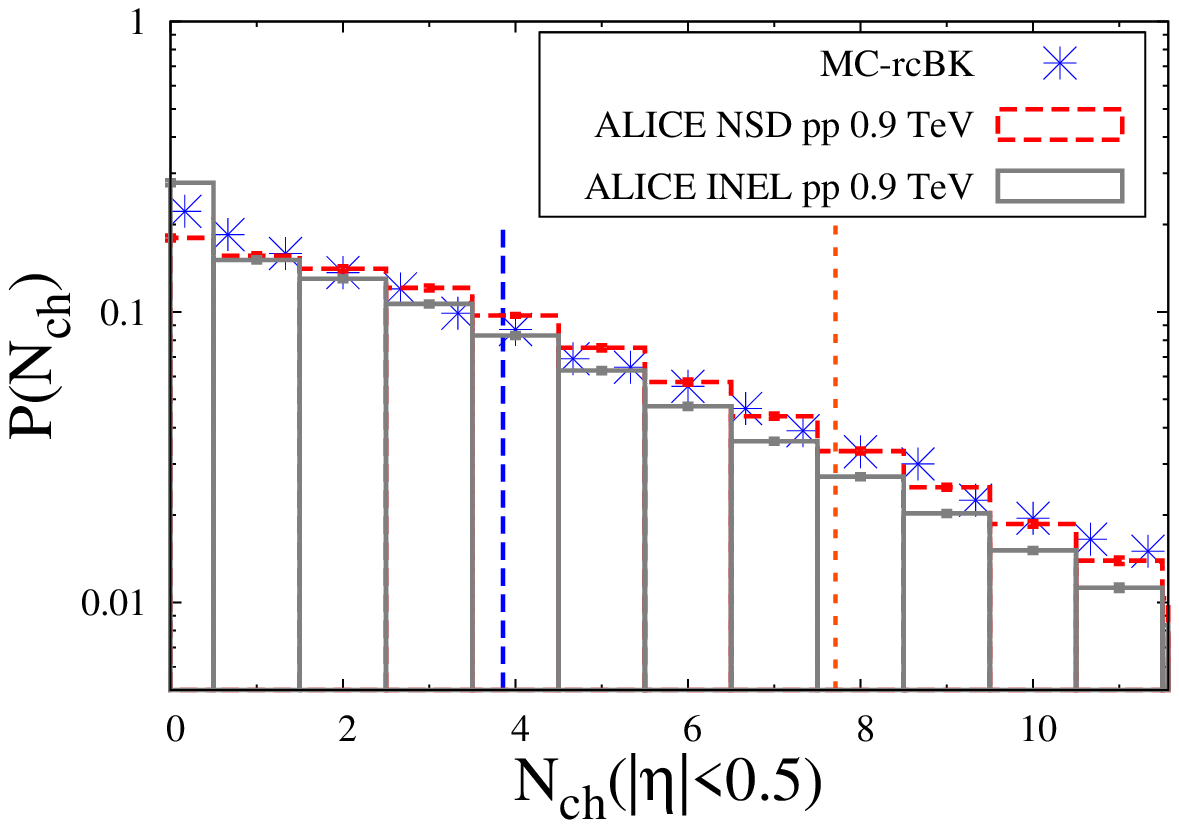}
\includegraphics[width=0.32\textwidth]{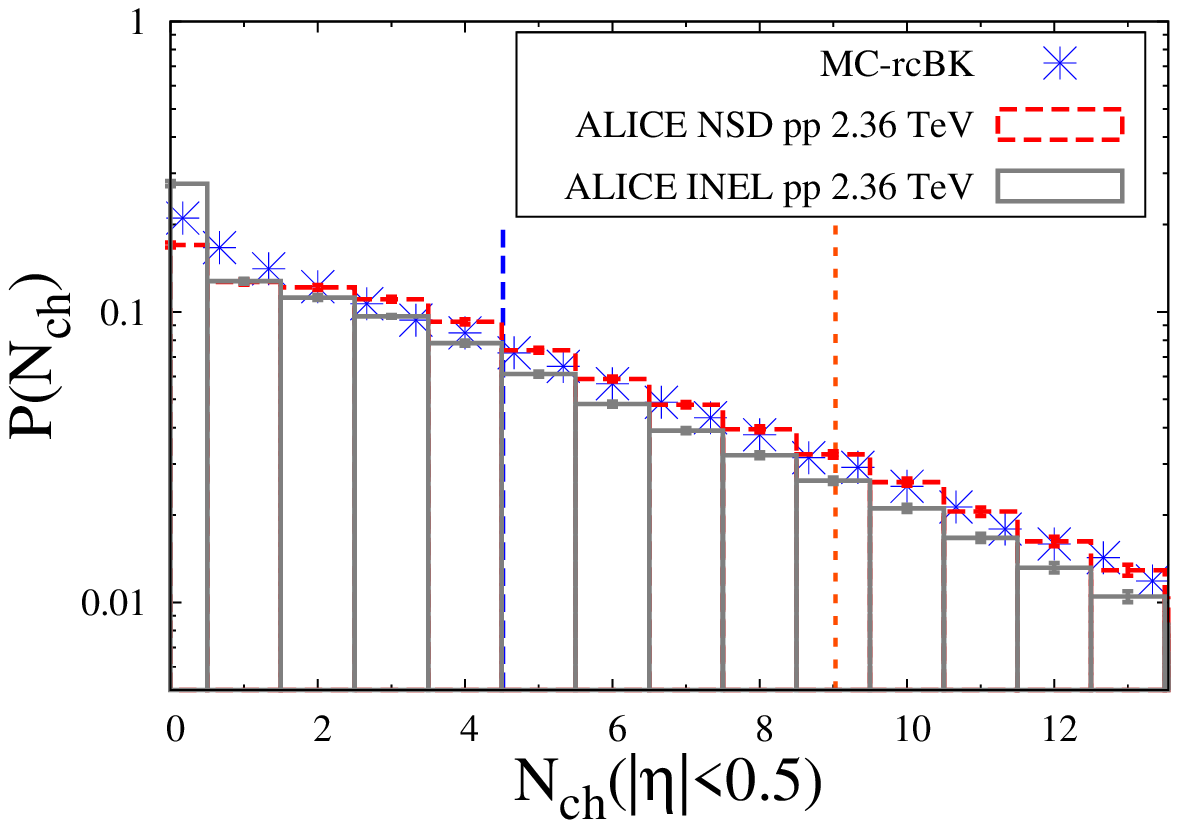}
\includegraphics[width=0.32\textwidth]{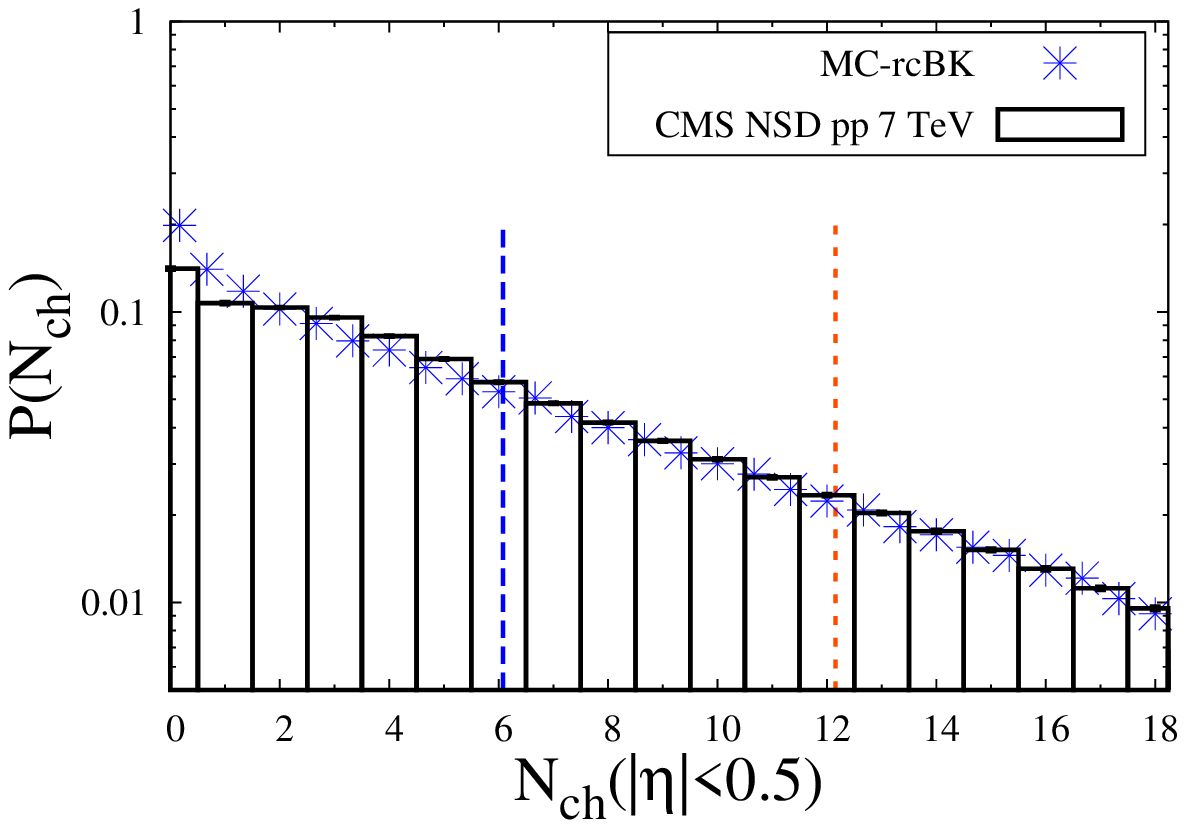}
\end{center}
\vspace*{-0.8cm}
\caption{\label{fig:Pn_pp}(Color online) 
{\bf Left}: multiplicity distribution of charged
  particles at $|\eta|<0.5$ in $pp$ collisions at $\sqrt{s}=900$~GeV.
Stars show the result of our calculation (see text) while solid and
dashed histograms correspond to data taken by the ALICE collaboration
with the ``NSD'' and ``INEL'' triggers,
respectively~\cite{Aamodt:2010ft}. The dashed vertical lines indicate
the average and two times the average multiplicity, respectively.
{\bf Center}: Same at $\sqrt{s}=2360$~GeV.
{\bf Right}: Same at $\sqrt{s}=7000$~GeV compared to CMS NSD
data~\cite{Khachatryan:2010nk}.
}
\end{figure}

We first analyze the multiplicity distributions in proton-proton
collisions at LHC energies (fig.~\ref{fig:Pn_pp}). We
concentrate on the bulk of the distributions, $N_{\rm ch}\lsim 3 \langle N_{\rm
  ch}\rangle$ where $\langle N_{\rm ch}\rangle$ denotes the average
charged particle multiplicity at a given energy. Over this range the
data can be described reasonably well by a NBD with constant
\be \label{eq:kpp}
k_{pp} = \frac{1}{\pi} \; \Delta^2 {\bf r}_\perp
\Delta\eta\; \Lambda^2_{\rm QCD}~.  
\ee 
Here $\Delta\eta=1$ and $\Delta^2 {\bf r}_\perp$ is the area of a cell
in transverse coordinate space over which we integrate
eq.~(\ref{eq:kt}). Also, we choose $\Lambda_{\rm QCD}=0.24$~GeV.
Numerically, $k/\bar n\simeq 0.16$ for $p+p$ collisions at 2.36~TeV.
We have checked that a weak energy dependence of $k$ is allowed as
long as it does not change the distribution $P(N_{\rm ch})$
appreciably over the range that we are interested in. The tails of
$P(N_{\rm ch})$ could be more sensitive to the detailed dependence of
$k$ on energy\footnote{The same applies to rapidity intervals bigger
  than $|\eta|<0.5$; UA5 found that $k$ then actually {\em decreases}
  with energy~\cite{UA5}.} but we do not explore the region $N_{\rm
  ch}>3\langle N_{\rm ch}\rangle$ here; see, for example,
ref.~\cite{Tribedy:2010ab}.

The most important consequence from~(\ref{eq:kpp}) is that since
$k$=const and smaller than the average multiplicity $\bar n$, it
follows that our multiplicity distributions satisfy
Koba-Nielsen-Olesen (KNO) scaling~\cite{Koba:1972ng}. That is, the
probability distribution $P(N_{\rm ch})$ is independent of energy if
expressed in terms of $z\equiv N_{\rm ch} / \langle N_{\rm
  ch}\rangle$; for $\bar n\gg k$ and in the region $z>k/\bar n$ the
NBD~(\ref{eq:NBD}) can be written in the form of a Gamma distribution
\be \label{eq:Gamma}
\bar n \, P(n) \, dz  \sim z^{k-1} e^{- kz}\, dz~.
\ee
We show the KNO scaling function in $pp$
collisions explicitly in fig.~\ref{fig:Pn_KNO} below.

\begin{figure}[htb]
\begin{center}
\includegraphics[width=0.45\textwidth]{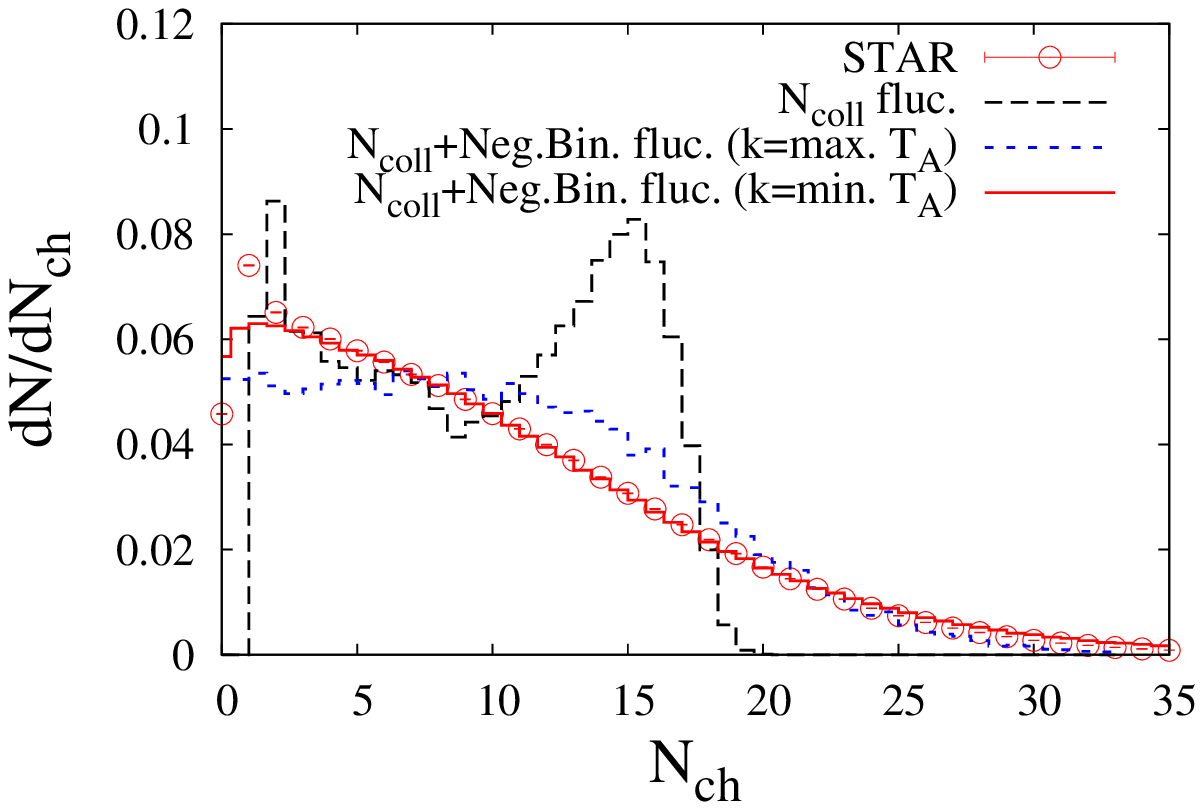}
\includegraphics[width=0.45\textwidth]{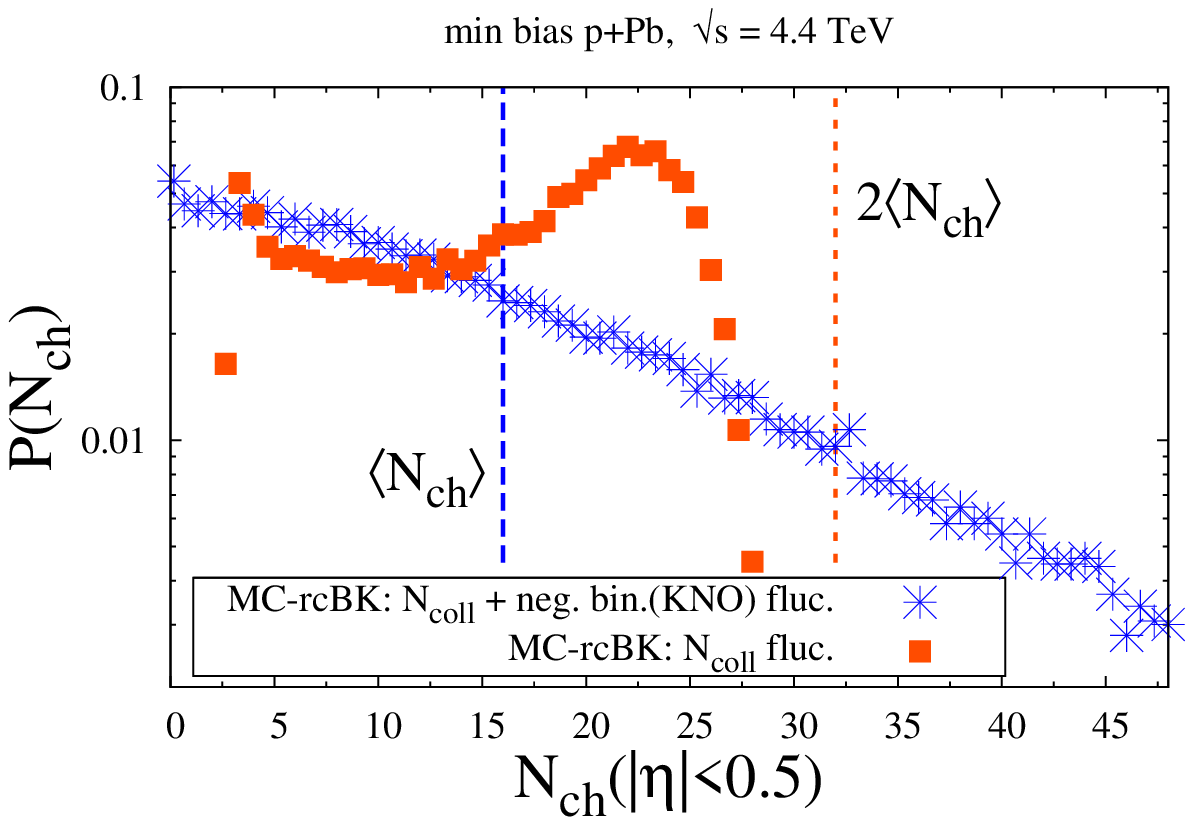}
\end{center}
\vspace*{-0.8cm}
\caption{\label{fig:Pn_pA}(Color online) 
{\bf Left}: multiplicity distribution of
  charged particles at $|\eta|<0.5$ in min.\ bias $d+Au$ collisions at
  $\sqrt{s}=200$~GeV.  Various models (see text) are compared to
  uncorrected data from the STAR collaboration (circles)
  ~\cite{STAR:2008ez}. {\bf Right}: multiplicity distribution
  predicted for min.\ bias $p+Pb$ collisions at $\sqrt{s}=4400$~GeV.
}
\end{figure}
In fig.~\ref{fig:Pn_pA} we compare the calculated charged particle
multiplicity distribution in $d+Au$ collisions at $\sqrt{s}=200$~GeV
to uncorrected data from STAR~\cite{STAR:2008ez}. As described above,
here we include also fluctuations of the number of participants
$N_{\rm part}$ and of the number of binary collisions $N_{\rm coll}$
which arise for different configurations of nucleons in the target
nucleus. Within our formalism, $N_{\rm coll}$ fluctuations alone are
insufficient to reproduce the experimental multiplicity
distribution. In this case we obtain a peak in $P(N_{\rm ch})$ before
the cutoff of the distribution which can be traced back to the fact
that $N_{\rm ch}$ does not increase linearly with the density of
sources when the latter is high. This ``saturation'' of particle
production is also responsible for the higher elliptic eccentricity of the
collision zone than obtained from simple linear estimates~\cite{CGCecc}.

Additional intrinsic fluctuations with
\be
k_{\rm d+Au} = k_{pp} \cdot \mathrm{min}\left(T_A({\bf r}_\perp),
T_B({\bf r}_\perp)\right)\, \sigma_0
\ee
lead to a good fit to the data; such scaling of $k$ with the number
of sources is expected due to the way that negative binomial
distributions add\footnote{If $x$ and $y$ are two random variables
  with a negative binomial distribution with mean $\mu$ and
  fluctuation parameter $k$ then $z=x+y$ also follows a negative
  binomial distribution with mean $2\mu$ and $k_z=2k$. Hence $k$ is an
  extensive quantity proportional to volume, just
as $\bar n$.}. On the other
hand, $k_{\rm d+Au}\sim \mathrm{max}\left(T_A({\bf r}_\perp), T_B({\bf
    r}_\perp)\right)$ produces a multiplicity distribution inbetween
the above cases, exhibiting too little fluctuations. Once again, it is
reasonable that the magnitude of fluctuations is determined mostly by
the dilute source (as also assumed in ref~\cite{Tribedy:2010ab}).
Our prediction for $p+Pb$ collisions at LHC is shown in
fig.~\ref{fig:Pn_pA} on the right; this corresponds to $\langle N_{\rm
  ch}\rangle \simeq 16$ and $k$ from eq.~(\ref{eq:kpp}). (A prediction for
the multiplicity distribution in $p+Pb$ collisions at the LHC from the
``KLN model'' was shown previously in ref.~\cite{DKLN}).

\begin{figure}[htb]
\begin{center}
\includegraphics[width=0.6\textwidth]{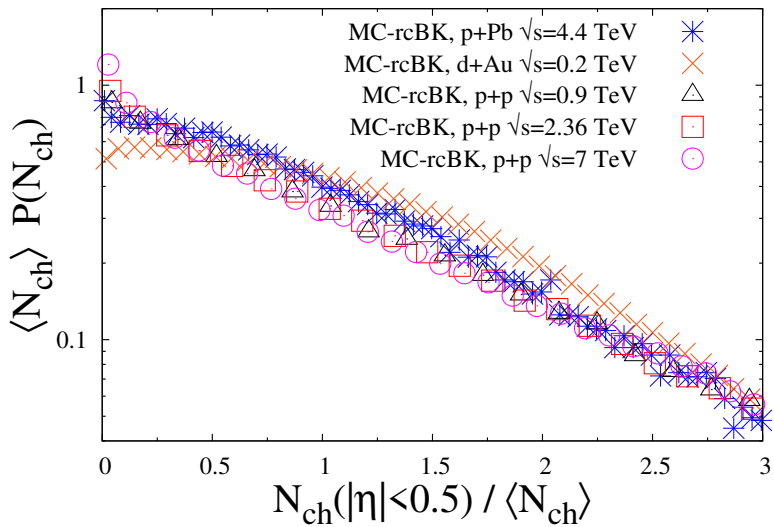}
\end{center}
\vspace*{-0.8cm}
\caption{\label{fig:Pn_KNO}(Color online)
KNO scaling plot of the multiplicity distributions
  of charged particles at $|\eta|<0.5$.
}
\end{figure}
Due to the presence of $N_{\rm coll}$ fluctuations our multiplicity
distribution for $p+Pb$ does not exhibit exact KNO scaling, as seen in
fig.~\ref{fig:Pn_KNO}. Nevertheless, for $|\eta|<0.5$ and $N_{\rm ch} \lsim 3
\langle N_{\rm ch}\rangle$ we predict relatively small deviations from the KNO
scaling function determined from $p+p$ collisions. This is an
important check for the presence of strong intrinsic particle
production fluctuations (at fixed $N_{\rm part}$ and $N_{\rm coll}$)
for a heavy-ion target.

We now proceed to discuss the relevance of particle production
fluctuations for various harmonic moments of the ``eccentricity'' of
gluons produced in the initial state of heavy-ion collisions. We
define moments of the initial density distribution (preceding the
hydrodynamic expansion in $A+A$ collisions) in terms of the
eccentricities~\cite{Alver:2010gr,Sorensen:2011hm}
\begin{equation}
\epsilon_n = \frac{\sqrt{\langle r^2\cos n\phi\rangle^2 +
\langle r^2\sin n\phi\rangle^2}}{\langle r^2\rangle} ~.
\end{equation}
Other definitions are sometimes also used in the literature, see for
example~\cite{Petersen:2010cw,Teaney:2010vd,Gardim:2011xv}.
$\langle\cdot\rangle$ denotes an average over the distribution of
produced gluons in the transverse plane, $dN/d\eta d^2{\bf r}_\perp$;
and ${\bf r}_\perp = r (\cos\phi,\sin\phi)$.

The
eccentricities $\epsilon_n$ are of interest because through
hydrodynamic response they generate the flow harmonics and angular
correlations in the final state of heavy-ion
collisions~\cite{hydro_vn,Alver:2010gr,Sorensen:2011hm,Petersen:2010cw,Teaney:2010vd,Gardim:2011xv,Qiu:2011hf,Schenke:2011bn,Lacey:2010hw,deSouza:2011rp,Werner:2010aa,Xu:2011jm,Ma:2010dv}.
(Fluctuations in small-$x$ evolution may also lead to detectable
azimuthal momentum anisotropies in high-multiplicity $pp$ collisions
at the LHC~\cite{Avsar:2010rf} which are not due to ``flow''). Flow
harmonics in heavy-ion collisions have been published by the
PHENIX~\cite{Adare:2011tg} and ALICE~\cite{Aamodt:2011by}
collaborations.

\begin{figure}[htb]
\begin{center}
\includegraphics[width=0.45\textwidth]{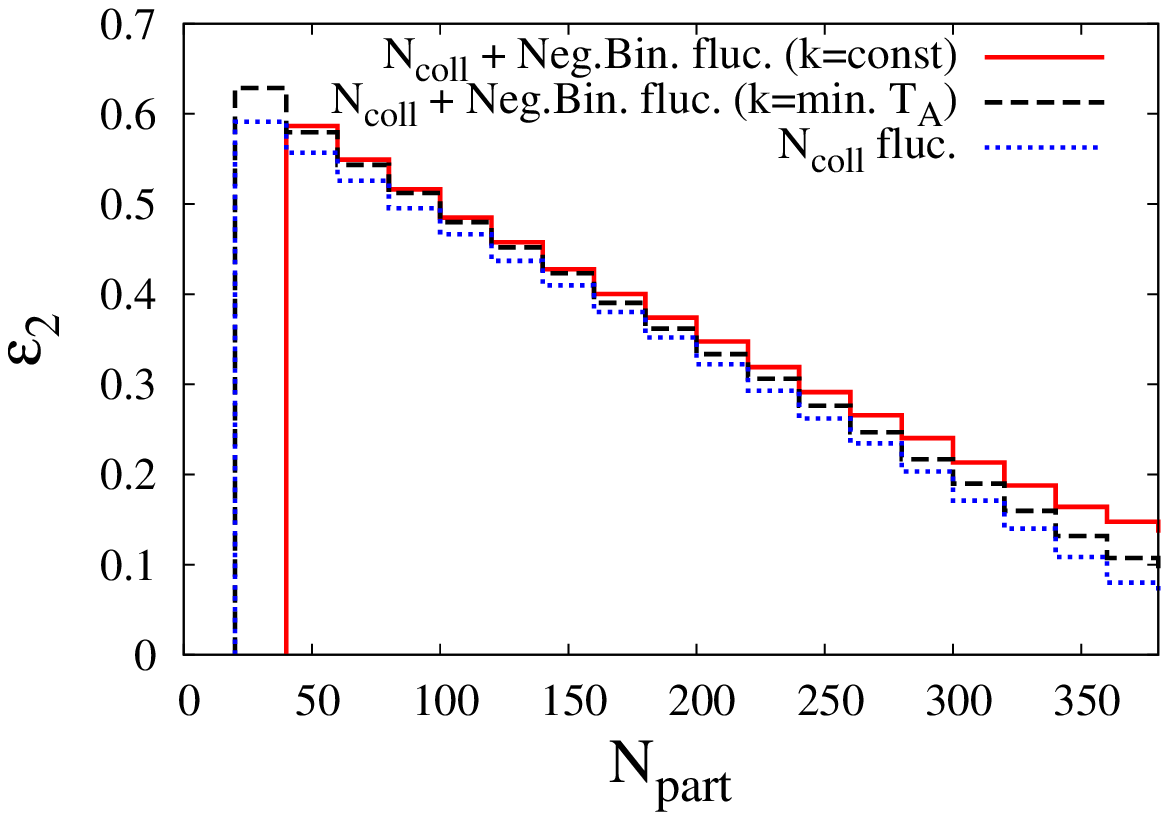}
\includegraphics[width=0.45\textwidth]{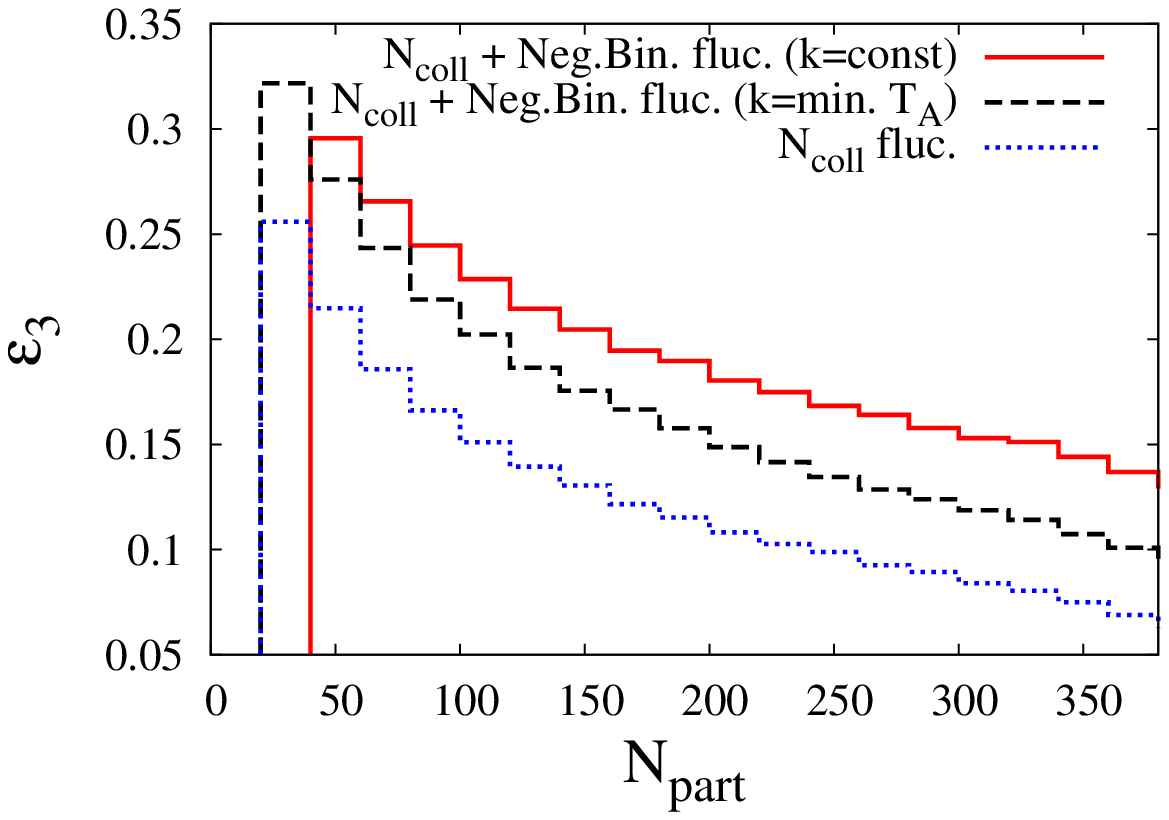}
\includegraphics[width=0.45\textwidth]{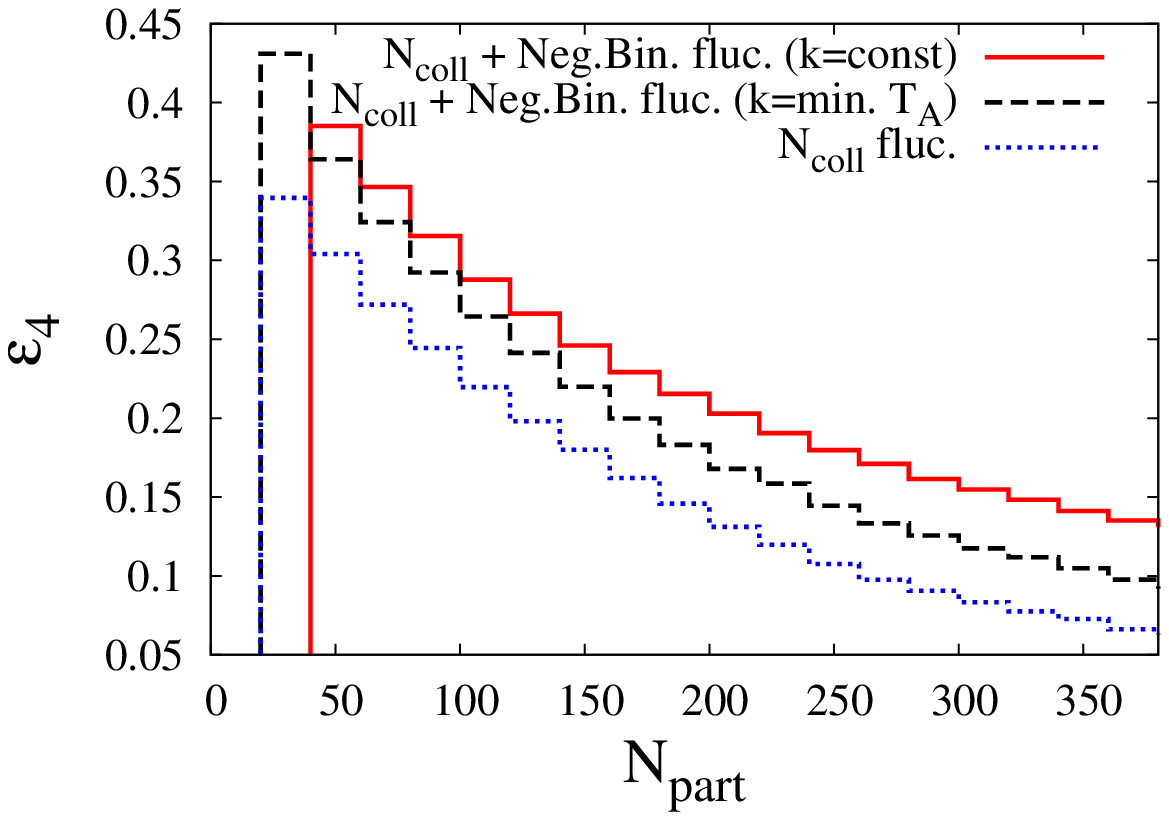}
\includegraphics[width=0.45\textwidth]{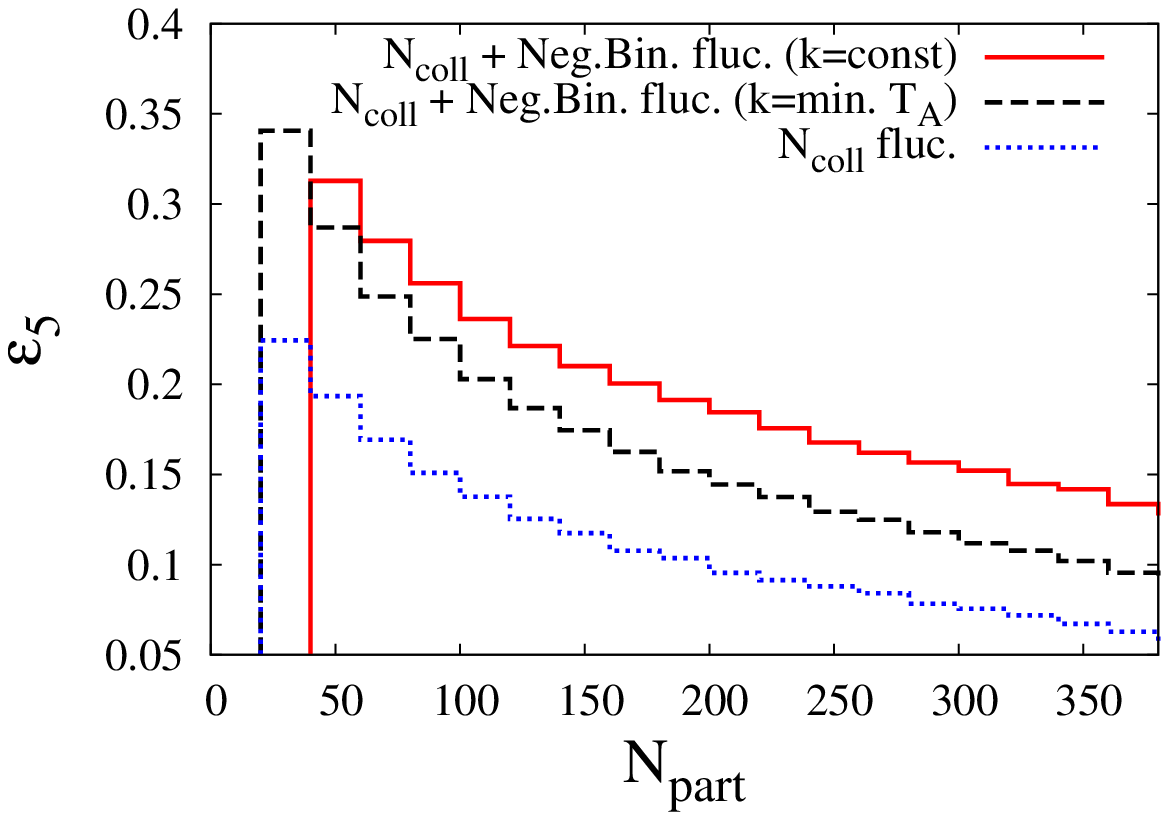}
\end{center}
\vspace*{-0.7cm}
\caption{\label{fig:eps_n}(Color online) 
Centrality dependence of various moments
  $\epsilon_n$ of the eccentricity in $Au+Au$ collisions at
  $\sqrt{s}=200$~GeV. Dotted lines correspond to local fluctuations of
  $N_{\rm coll}$ (``geometry fluctuations'') only; dashed and full
  lines add particle production fluctuations according to a negative
  binomial distribution with $k=k_{pp}=\;$const or
  $k\sim\mathrm{min}(T_A,T_B)$, respectively.
}
\end{figure}
In fig.~\ref{fig:eps_n} we compare the centrality dependence of
$\epsilon_2$ -- $\epsilon_5$ for three different models. In all cases,
the lowest curve corresponds to the model with ``geometry
fluctuations'' only, as usually considered in the literature. If
particle production fluctuations according to a negative binomial
distribution are added then in general $\epsilon_n$ increases. The
effect on $\epsilon_2$ at $N_{\rm part}\lsim250$ is small since the
ellipticity of the overlap zone is of course dominated by the ``almond
shaped'' geometry of heavy-ion collisions at finite impact parameter.
On the other hand, we observe a large increase of $\epsilon_n$ for all
$n\ge3$ over the entire range of impact parameters, as well as of
$\epsilon_2$ for $N_{\rm part}\to 2A$, since these observables are
fluctuation dominated. Most importantly, the ratio
$\epsilon_3/\epsilon_2$ in mid-central collisions increases
significantly. The largest increase is obtained if the fluctuation
parameter $k=k_{pp}$ does not increase with the density of sources.
Even for the more realistic case where $k\sim\mathrm{min}(T_A,T_B)$,
higher-order eccentricities can increase by as much as 50\%. We
mention also that simulations using the DIPSY Monte-Carlo which
performs the small-$x$ dipole evolution stochastically have predicted
a large $\epsilon_3$~\cite{DIPSY}, although the relation to KNO
scaling in $pp$ and $pA$ collisions at the LHC had not been pointed out.

To summarize our main results: we found that in order to reproduce the
measured multiplicity distribution in $d+Au$ collisions at RHIC within
the CGC approach it is important to take into account particle
production fluctuations (according to a negative binomial
distribution). We predict that these dominate over Glauber
fluctuations also for $p+Pb$ collisions at the LHC, resulting in a
multiplicity distribution which is close to the KNO scaling function
measured in $p+p$ collisions. The effect of particle production
fluctuations can be large also for some observables in heavy-ion
collisions, such as for higher-order eccentricities. It will be
interesting to see how this reflects in higher-order flow coefficients
predicted by viscous hydrodynamics or in the centrality dependence of
the jet quenching parameter $R_{AA}(p_\perp)$~\cite{Betz:2012qq}.

\section*{Acknowledgements}
We gratefully acknowledge the kind hospitality of the Institute of
Physics at the University of Tokyo;  our stay
was supported by grant KAKENHI(22340064).
A.D.\ also acknowledges support by the DOE Office of
Nuclear Physics through Grant No.\ DE-FG02-09ER41620; and from The
City University of New York through PSC-CUNY Research
grant 64132-00~42. The work of Y.N.\ was partly supported by
Grant-in-Aid for Scientific Research No.~20540276.

\appendix
\section{Eccentricities for heavy-ion collisions at LHC energies}
\begin{figure}[htb]
\begin{center}
\includegraphics[width=0.45\textwidth]{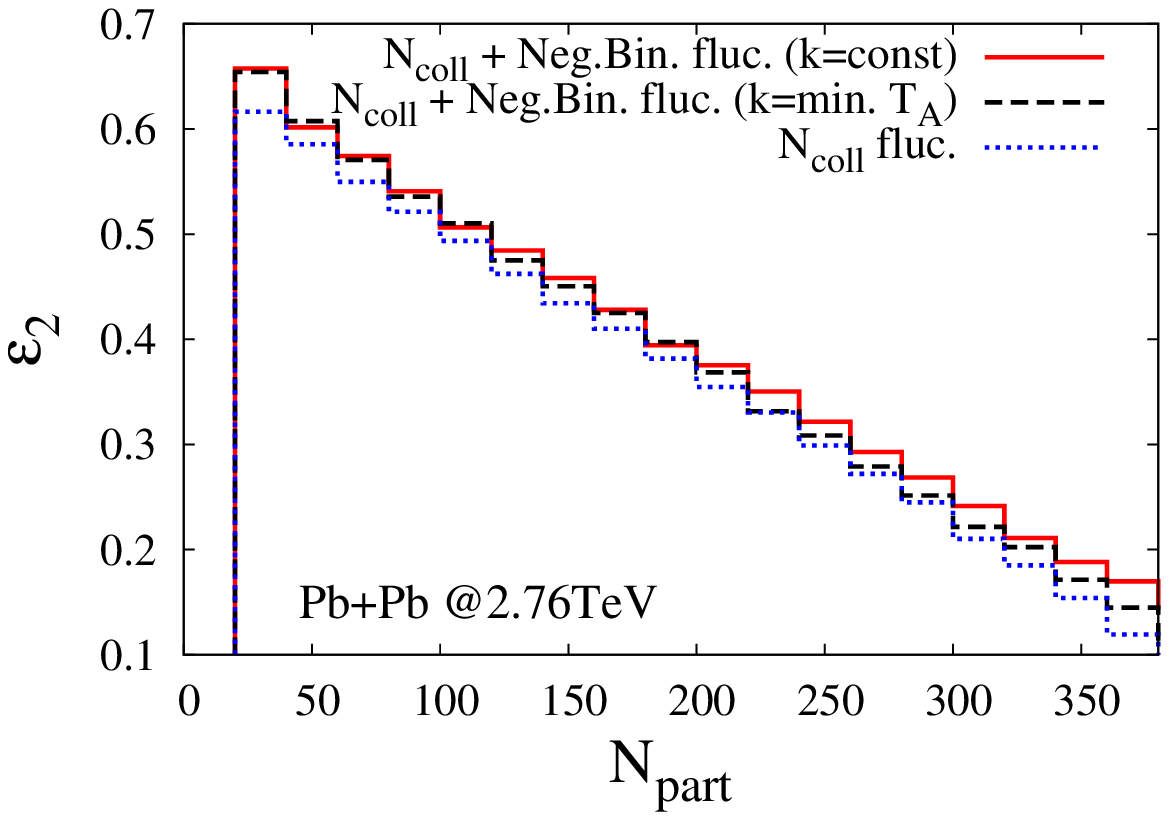}
\includegraphics[width=0.45\textwidth]{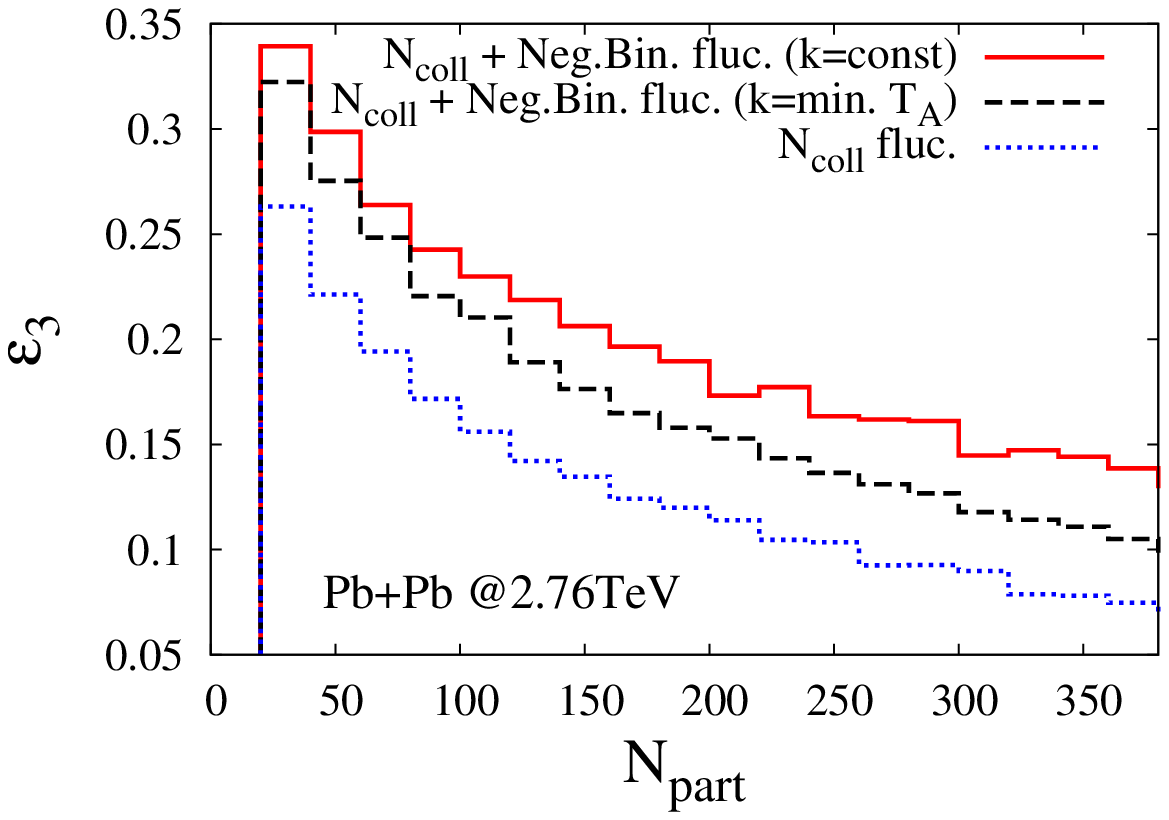}
\includegraphics[width=0.45\textwidth]{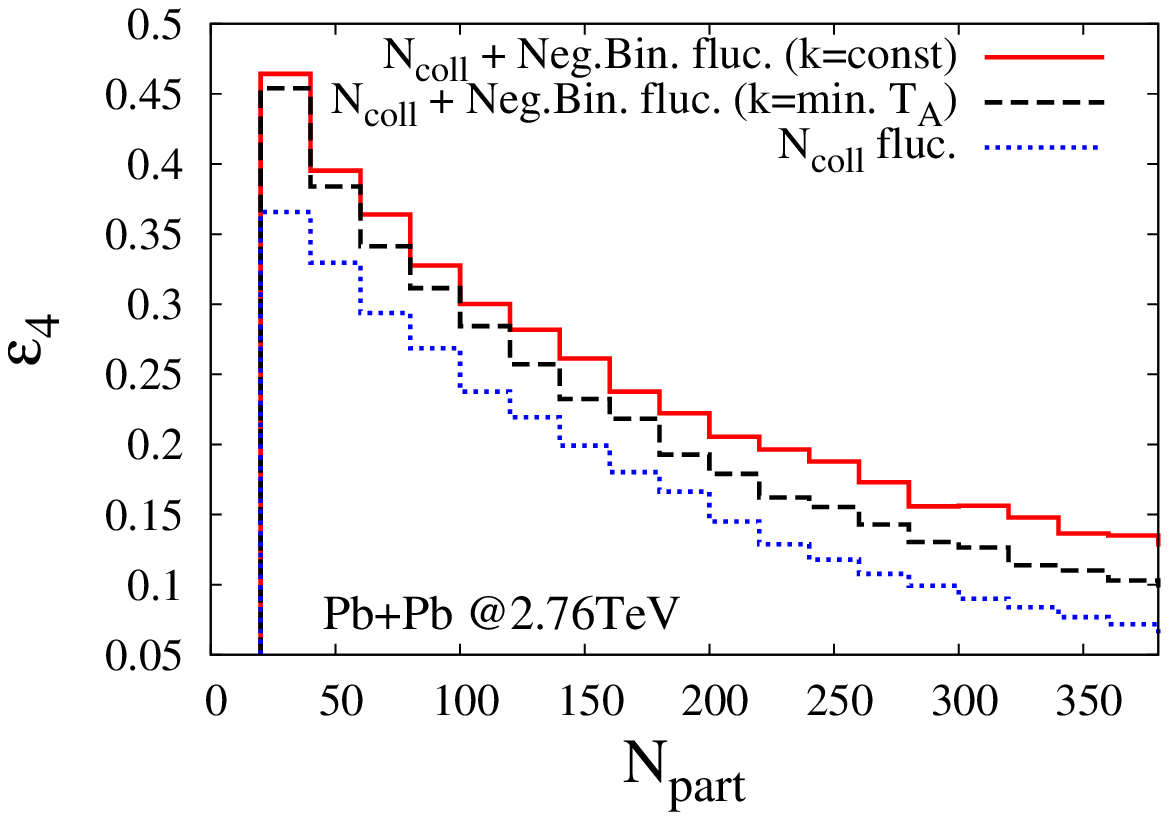}
\includegraphics[width=0.45\textwidth]{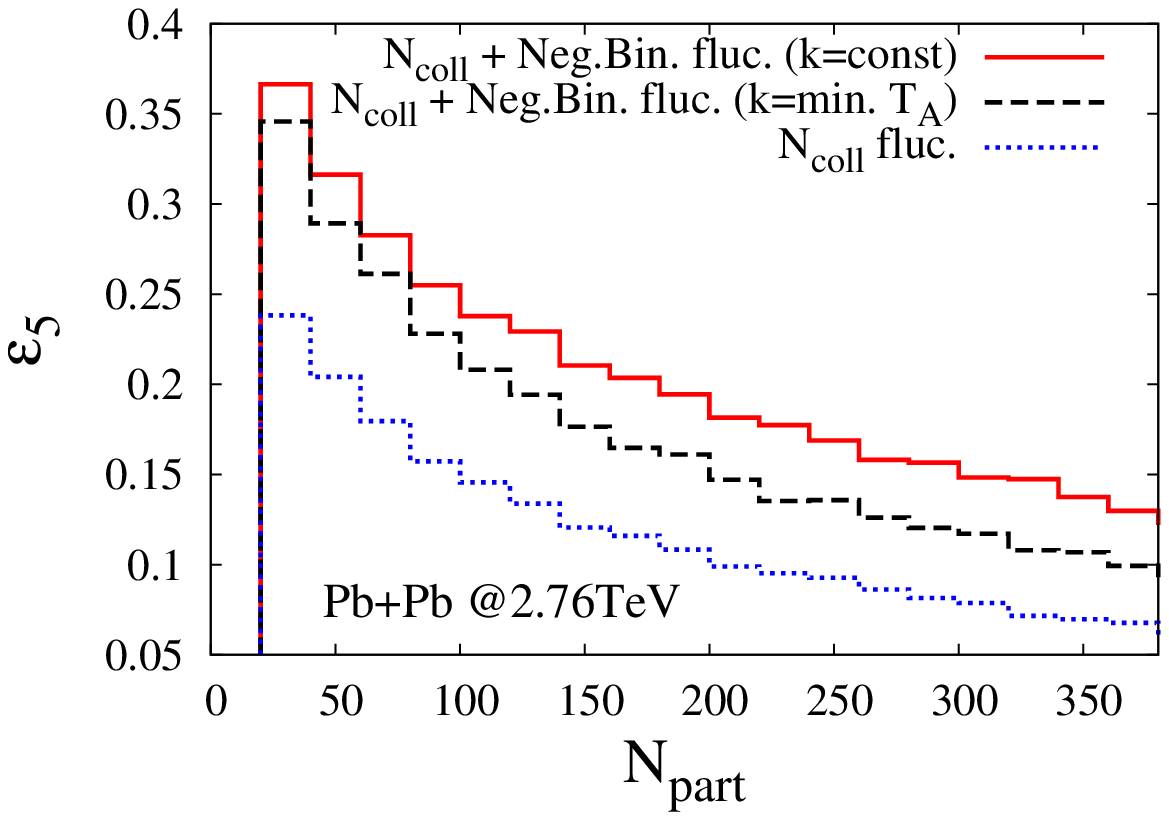}
\end{center}
\vspace*{-0.7cm}
\caption{\label{fig:LHCeps_n}(Color online)
  Centrality dependence of various moments
  $\epsilon_n$ of the eccentricity in $Pb+Pb$ collisions at
  $\sqrt{s}=2.76$~TeV. Dotted lines correspond to local fluctuations of
  $N_{\rm coll}$ (``geometry fluctuations'') only; dashed and full
  lines add particle production fluctuations according to a negative
  binomial distribution with $k=k_{pp}=\;$const or
  $k\sim\mathrm{min}(T_A,T_B)$, respectively.
}
\end{figure}

\end{document}